\documentclass[twocolumn,aps,prb,english]{revtex4-2}

\usepackage[colorlinks=true,allcolors=blue]{hyperref}
\usepackage{array}
\usepackage{amsmath}
\usepackage{amssymb}
\usepackage{graphicx}
\usepackage{babel}
\usepackage{mathrsfs}
\usepackage{amsfonts}
\usepackage{epstopdf}
\usepackage{multirow}
\usepackage{color}
\usepackage{bm}
\usepackage{amsthm}

\usepackage{xcolor}
\usepackage{lipsum}
\usepackage{tikz,fp}
\usepackage{tikz-cd}
\usepackage[export]{adjustbox}
\usepackage{hhline}
\usepackage{tikz}
\usetikzlibrary{quantikz2}

\newcommand{\tr}{\mathrm{tr}}
\newcommand{\dg}{\dagger}

\newcommand{\hf}{\frac{1}{2}}

\begin{document}

\title{Numerical investigations of the extensive entanglement Hamiltonian in quantum spin ladders}
\author{Chengshu Li}
\email{lichengshu272@gmail.com}
\author{Xingyu Li}
\author{Yi-Neng Zhou}
\affiliation{Institute for Advanced Study, Tsinghua University, Beijing, 100084, China}
\date{\today}

\begin{abstract}
Entanglement constitutes one of the key concepts in quantum mechanics and serves as an indispensable tool in the understanding of quantum many-body systems. In this work, we perform extensive numerical investigations of extensive entanglement properties of coupled quantum spin chains. This setup has proven useful for e.g. extending the Lieb--Schultz--Mattis theorem to open systems, and contrasts the majority of previous research where the entanglement cut has one lower dimension than the system. We focus on the cases where the entanglement Hamiltonian is either gapless or exhibits spontaneous symmetry breaking behavior. We further employ conformal field theoretical formulae to identify the universal behavior in the former case. The results in our work can serve as a paradigmatic starting point for more systematic exploration of the largely uncharted physics of extensive entanglement, both analytical and numerical.
\end{abstract}

\maketitle

\section{Introduction}
Over the past few decades, quantum entanglement has firmly established its status as an indispensable concept in the understanding of quantum many-body systems. The study of (topological) entanglement entropy was pioneered by Levin--Wen~\cite{Levin2006} and Kitaev--Preskill \cite{Kitaev2006}, where the subleading term in the entanglement entropy scaling diagnoses the topological order \cite{Jiang2012}, followed by the proposal of Li--Haldane \cite{Li2008} to explore the entanglement Hamiltonian $K=-\log \rho$ as a finer probe, the spectrum of which is identified with the (physical) edge spectrum \cite{Fidkowski2010,Pollmann2010,Qi2012}. Research efforts along these lines have significantly enriched our understanding of quantum many-body physics \cite{Zeng2019}, in particular, how interesting physics can be extracted from a single ground state wavefunction using entanglement tools \cite{Amico2008,Jiang2012,Zhang2012,Cincio2013,Kim2022a,Kim2022b,Zou2022,Fan2022,Fan2023,Liu2023}. Recently, simulation and measurement of entanglement Hamiltonian have also gained traction on the experimental frontier \cite{Dalmonte2018,Choo2018,Zache2022,Dalmonte2022}. It is interesting to note that related results have been developed in parallel in high energy physics, largely in a decoupled way from the quantum many-body community~\cite{Bisognano1975,Bisognano1976,Witten2018,Dalmonte2022}.

\begin{figure}
    \centering
    \includegraphics[width=0.47\textwidth]{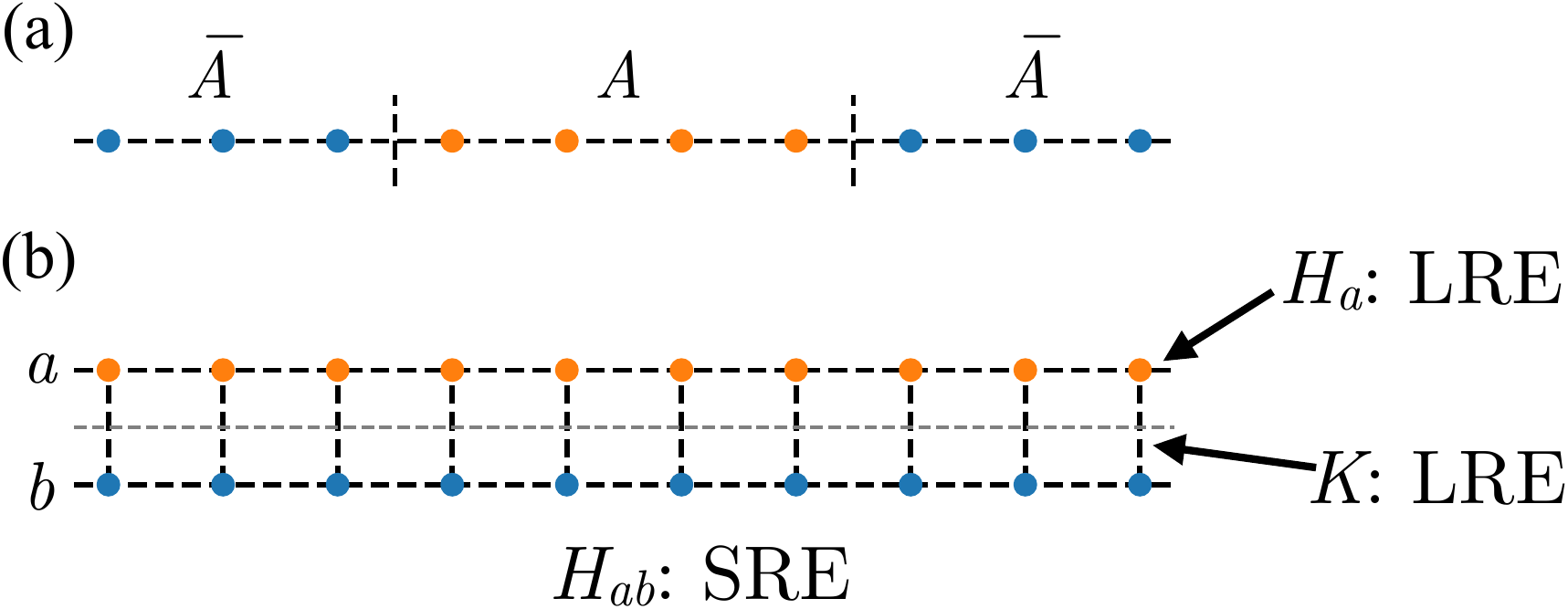}
    \caption{Sub-dimensional vs. extensive entanglement. (a) A zero-dimensional cut of a one-dimensional system. This, together with higher-dimensional analogies, is the scenario that the majority of research efforts have focused on. (b) A coupled spin chain and its extensive entanglement cut. In this work, we focus on the case where the chain-$a$ has a long-range entangled (LRE), translation and rotation invariant Hamiltonian, while the total ladder comprising the chain-$a$ and the chain-$b$ is short-range entangled (SRE) with a symmetry-preserving coupling. We can view the chain-$a$ as the system, and the chain-$b$ as the bath. Under these conditions one expects that the extensive entanglement Hamiltonian of the chain-$a$ defined by an extensive entanglement cut, depicted as the gray dashed line, is also LRE.}
    \label{fig:1}
\end{figure}

The majority of works on the entanglement entropy and entanglement spectrum are framed in a setup where the entanglement cut has one dimension lower than the system. For instance, both the aforementioned proposals of topological entanglement entropy and entanglement spectrum deal with a one-dimensional (1d) cut in a two-dimensional (2d) system. Nevertheless, much remains unknown for the case where the entanglement cut has the \emph{same} dimension as the system, sometimes referred to as ``extensive'' or ``bulk'' entanglement \cite{Poilblanc2010, schliemann2012entanglement,rao2014criticalentanglement,Hsieh2014,Vijay2015,Santos2015,gu2016holographic,nehra2018manybody,Zhu2019,Li2023}. This dichotomy is illustrated in Fig.~\ref{fig:1}. Entanglement spectra in these setups were shown to be connected to the edge low-energy spectrum, thereby extending the Li-Haldane conjecture\cite{Poilblanc2010, schliemann2012entanglement}. The extensive entanglement is recently brought into focus by Ref.~\cite{zhou2023} where the entanglement Hamiltonian is proposed as a natural setting to revive the Lieb--Schultz--Mattis (LSM) theorem~\cite{Lieb1961,Oshikawa2000,Hastings2004} where an LSM chain with spin-1/2 onsite Hilbert space and spin-rotational and translational symmetry is symmetrically coupled to a bath and becomes short-ranged correlated. 

\begin{figure*}
    \centering
    \includegraphics[width=0.9\textwidth]{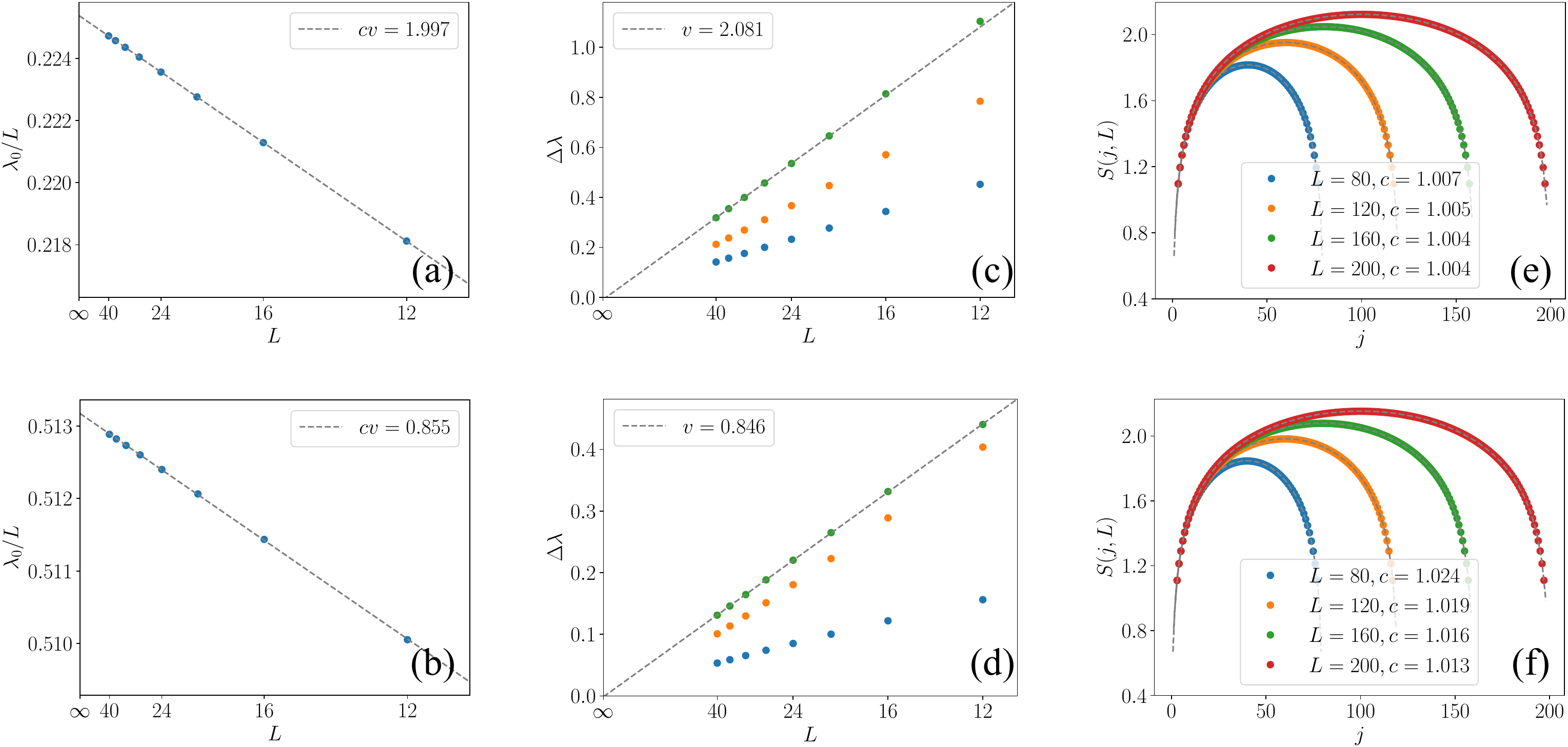}
    \caption{The numerical results for the AKLT ladders, with (a, c, e) for Model I and (b, d, f) for Model II. (a, b) Scaling of the lowest entanglement energy $\lambda_0$ using Eq.~\eqref{eq:cv}, giving the product of the central charge $c$ and the velocity $v$. (c, d) Scaling of the entanglement energy gap $\Delta\lambda$ using Eq.~\eqref{eq:v}, giving $v$. These together give an estimate of $c=0.96$ for Model I and $c=1.01$ for Model II. 
    Here, we use the gap between $\lambda_0$ and $\lambda_3$ (green dots) to extract $v$ as $\lambda_3$ corresponds to the first excited state in the conformal tower. This is identified from small-size exact diagonalization results where its lattice momentum is found to be $2\pi/L$  (not shown). The blue and orange dots correspond to states at momentum $\pi$ and belong to other conformal towers. 
    (e, f) Fitting of the entanglement entropy using the Calabrese--Cardy formula, Eq.~\eqref{eq:cc}. All the calculations are carried out with periodic boundary conditions and bond dimension $\chi=1000$.}
    \label{fig:aklt}
\end{figure*}

Motivated by the results in Ref.~\cite{zhou2023}, in this work, we perform extensive numerical calculations of the entanglement properties in quantum spin ladders. In particular, we consider ladders where one leg satisfies the LSM conditions, i.e. having spin-1/2 onsite Hilbert space and spin-rotational and translational symmetry, while the ladder as a total has a unique gapped ground state. Phrased in a more modern language, the leg without coupling has a long-range entangled (LRE) ground state, while the ground state of the ladder is short-range entangled (SRE) \cite{Gioia2022}. We wish to explore whether the entanglement Hamiltonian again admits LRE, as dictated by the open system LSM theorem, see Fig.~\ref{fig:1}(b). In this paper, we consider the two models that were introduced in Ref.~\cite{zhou2023} and two more related ones --- two Affleck--Kennedy--Lieb--Tasaki (AKLT) ladders and two decohered Majumdar--Ghosh (MG) ladders. These are prototypical scenarios where the open system LSM theorem gives nontrivial predictions, and serve as a natural starting point for more systematic investigations along these lines.  In the former two with gapless entanglement spectra, we further use various conformal field theoretical (CFT) formulae to identify the corresponding low-energy theory. In all calculations, we perform extensive density-matrix renormalization group (DMRG) calculations to reach system sizes much larger than previous exact diagonalization results. Apart from lending full support to the open system LSM theorem with much greater details, our numerical results show that DMRG methods can be fruitfully applied to (generally only quasi-local) entanglement Hamiltonian, paving way for future studies in this direction. It is also notable that LSM systems can sometimes be viewed as boundaries of symmetry protected topological (SPT) phases \cite{Cho2017}, and our results can offer insights into these scenarios as well.

\begin{figure*}
    \centering
    \includegraphics[width=0.9\textwidth]{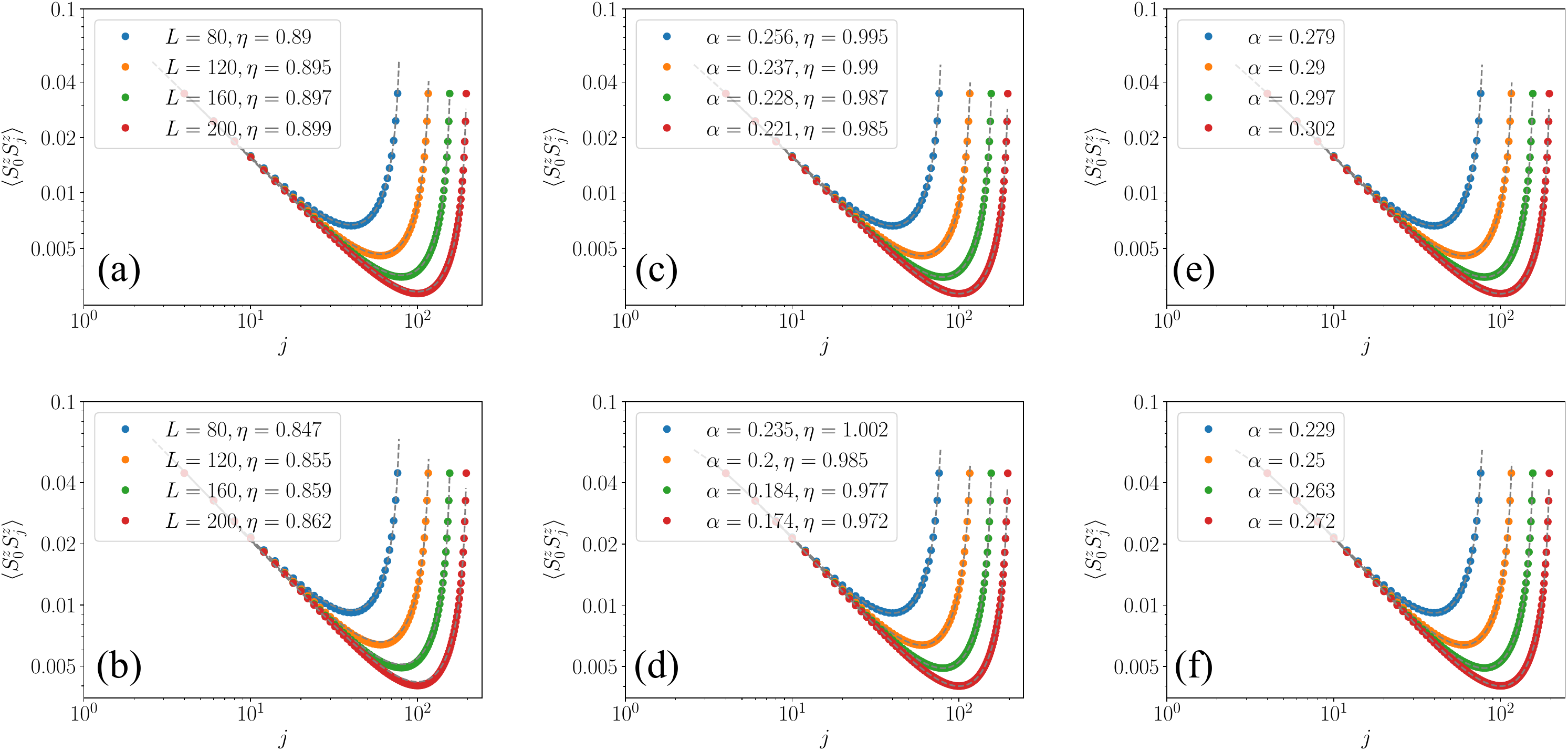}
    \caption{The two-point correlation function of the AKLT ladders, with (a, c, e) for Model I and (b, d, f) for Model II. We use three different fitting ans\"atze, Eqs.~(\ref{eq:eta}, \ref{eq:alpha_eta}, \ref{eq:alpha}), to fit the numerical data, corresponding to the left, middle and right columns. Overall, the ones with the logarithmic correction (c--f) fit better than those without (a, b). Fittings without the logarithmic correction (a, b) give the critical exponent $\eta$ around $0.9$, while those with both $\eta$ and the logarithmic exponent $\alpha$ as fitting parameters give $\eta\simeq1.0$. In (e, f) we fix $\eta=1$ and fit against $\alpha$, giving $\alpha\simeq0.3$. All the calculations are carried out with periodic boundary conditions and bond dimension $\chi=1000$.}
    \label{fig:aklt2}    
\end{figure*}

\section{Gapless entanglement Hamiltonian of the AKLT ladders}
We now focus on two AKLT ladders, the second of which was first proposed in Ref.~\cite{zhou2023}. Both models have the virtue that the ground state wavefunction can be written down in terms of an exact matrix product state (MPS). From this, the reduced density matrix can be formulated as an exact matrix product density operator (MPDO), without involving any approximations. The Hamiltonians of the two models both read
\begin{equation}
{H}_\mathrm{I,II}=\sum_{i=1}^L J_1\big({\bf {S}}_{i}\cdot {\bf {S}}_{i+1}+\frac{1}{3}({\bf {S}}_{i}\cdot {\bf {S}}_{i+1})^2\big)+J_2 {\bf {S}}_{i,a}\cdot {\bf {S}}_{i,b},
\end{equation}
where ${\bf {S}}_{i}={\bf {S}}_{i,a}+{\bf {S}}_{i,b}$ and $J_1>0$. The chain-$a$ has spin-$1/2$, while the chain-$b$ has spin-$1/2$ (spin-$3/2$) in Model I (II). We further take $J_2<0$ ($J_2>0$) in Model I (II), which guarantees that each rung has spin-1, the required on-site spin for the AKLT construction~\cite{Affleck1987,Affleck1988b}. Under these conditions, the frustration-free ground state is independent of $J_{1,2}$. Now it is straightforward to write down the MPS for the ground state,
\begin{equation}
\begin{split}
&\psi_{\{\mu\}}=\tr \prod_j A_{\mu_{j,a}\mu_{j,b}}^{(j)},\\
&(A_{\mu_{j,a}\mu_{j,b}}^{(j)})_{\nu_j\nu_{j+1}}
\propto\sum_{\mu\nu} C_{S_a,S_b,1}^{\mu_{j,a},\mu_{j,b},\mu} C_{\hf,\hf,1}^{\nu_{j},\nu,\mu} i\sigma_{\nu,\nu_{j+1}}^y.
\end{split}
\end{equation}
Here $C_{\cdots}^{\cdots}$ denotes the Clebsch--Gordon coefficients, the matrix product and the trace are over the auxiliary indices $\{\nu\}$, $S_a=1/2$, and $S_b=1/2$ ($S_b=3/2$) for Model I (II) as above. The idea is to first fuse $S_a$ and $S_b$ to spin-1, then split it to two spin-$1/2$, and finally twist one of the spin-$1/2$ to obtain a spin singlet between two nearest neighbor sites\footnote{Note that $i\sigma_{\mu\nu}^y\propto C_{1/2,1/2,0}^{\mu,\nu,0}$.}. Graphically, we have
\begin{equation}
\includegraphics[width=0.3\textwidth,valign=c]{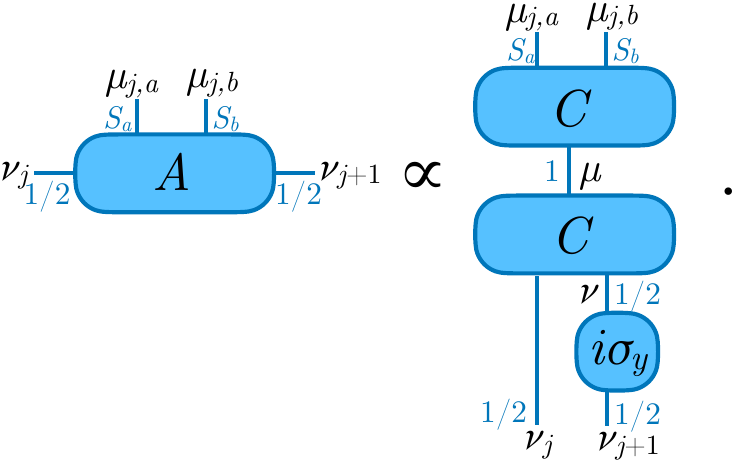}
\end{equation}
Tracing over $\{\mu_b\}$, the resultant MPDOs read
\begin{equation}
\begin{split}
&\rho_\mathrm{I,II}(L)=\tr M_\mathrm{I,II}^L, \\
&M_\mathrm{I}\propto\begin{pmatrix}
\frac{1}{2} & -S^- & -S^+ & 2P^\uparrow\\
S^+ & -\frac{1}{2} & 0 & S^+\\
S^- & 0 & -\frac{1}{2} & S^-\\
2P_\downarrow & -S^- & -S^+ & \frac{1}{2}
\end{pmatrix},\\
&M_\mathrm{II}\propto\begin{pmatrix}
1 & S^- & S^+ & 2-2S^z\\
-S^+ & -1 & 0 & -S^+\\
-S^- & 0 & -1 & -S^-\\
2+2S^z & S^- & S^+ & 1
\end{pmatrix},
\end{split}
\end{equation}
where we have used the standard spin-$1/2$ operators.

Before proceeding to the numerical results, we note a connection between our construction and integrable systems. Integrability has been identified in a previous work with a similar but distinct setup~\cite{Santos2015}. For rank-4 MPDO, Ref.~\cite{Katsura2015} studies a family of integrable models labeled by two parameters $x,y$~\footnote{We thank Hosho Katsura for bringing this to our attention.}. It is amusing to note that $M_\mathrm{I,II}$ can also be brought to the general form of Ref.~\cite{Katsura2015}, with $x=i/\sqrt{3},y=2/\sqrt{3}$ and $x=i/\sqrt{3},y=1/\sqrt{3}$ respectively. 
To see this, we use the unitary matrices 
\begin{equation}
U_\mathrm{I}=\frac{1}{\sqrt{2}}
\begin{pmatrix}
1 & 0 & 0 & -i\\
0 & -i & 1 & 0\\
0 & -i & -1 & 0\\
1 & 0 & 0 & i
\end{pmatrix},
\end{equation}
\begin{equation}
U_\mathrm{II}=\frac{1}{\sqrt{2}}
\begin{pmatrix}
1 & 0 & 0 & i\\
0 & i & -1 & 0\\
0 & i & 1 & 0\\
1 & 0 & 0 & -i
\end{pmatrix},
\end{equation}
to rotate the MPDO to the form in Ref.~\cite{Katsura2015} with $U_\alpha^\dg M_\alpha U_\alpha,\alpha=\mathrm{I,II}$.
While the original construction has both parameters $x,y$ real, in our case $x$ is complex. For our purpose, the numerical results can be deemed exact, and we will not further pursue the ramifications of integrability. We leave a full-fledged investigation along these lines for future research.

Equipped with an MPDO for $\rho$, we can apply the standard DMRG algorithm to find its eigenstates of the highest eigenvalues, corresponding to the lowest entanglement energy states of $K$. The eigenvalues $\lambda$ of $K$ are obtained by taking a logarithm of those of $\rho$. It is interesting to note that while some kind of locality is expected of $K$ \cite{zhou2023}, $\rho$ is a very non-local object, and yet DMRG works extremely well with its MPDO representation. The convergence is guaranteed from both very small truncation errors and a scaling of the bond dimension. The success of the algorithm can ultimately be attributed to the quasi-locality of $K$, which produces at most logarithmic entanglement growth in the one-dimensional case. Given the expectation of a gapless spectrum, we use the standard CFT formulae to fit the physical parameters. We start with the entanglement energy scaling, from which the entanglement velocity $v$ and the central charge $c$ can be extracted. We use~\cite{Francesco1997}
\begin{align}
&\frac{\lambda_0(L)}{L}=\lambda_\infty-\frac{\pi cv}{6L^2}+\cdots,\label{eq:cv}\\
&\Delta \lambda=\frac{2\pi v}{L}+\cdots,\label{eq:v}
\end{align}
where $\lambda_\infty$ is the entanglement energy density in the thermodynamic limit. The results for both models are shown in Fig.~\ref{fig:aklt}(a--d). Next, we partition the chain into two parts of $j$ and $L-j$ and fit the entanglement entropy with the Calabrese--Cardy formula~\cite{Calabrese2004},
\begin{equation}
S(j,L)=\frac{c}{3}\ln\left[\frac{L}{\pi}\sin\left(\frac{\pi j}{L}\right)\right]+S_{0}.\label{eq:cc}
\end{equation}
The results are plotted in Fig.~\ref{fig:aklt}(e, f). All these results are consistent with a $c=1$ CFT. 

\begin{figure*}
    \centering
    \includegraphics[width=0.9\textwidth]{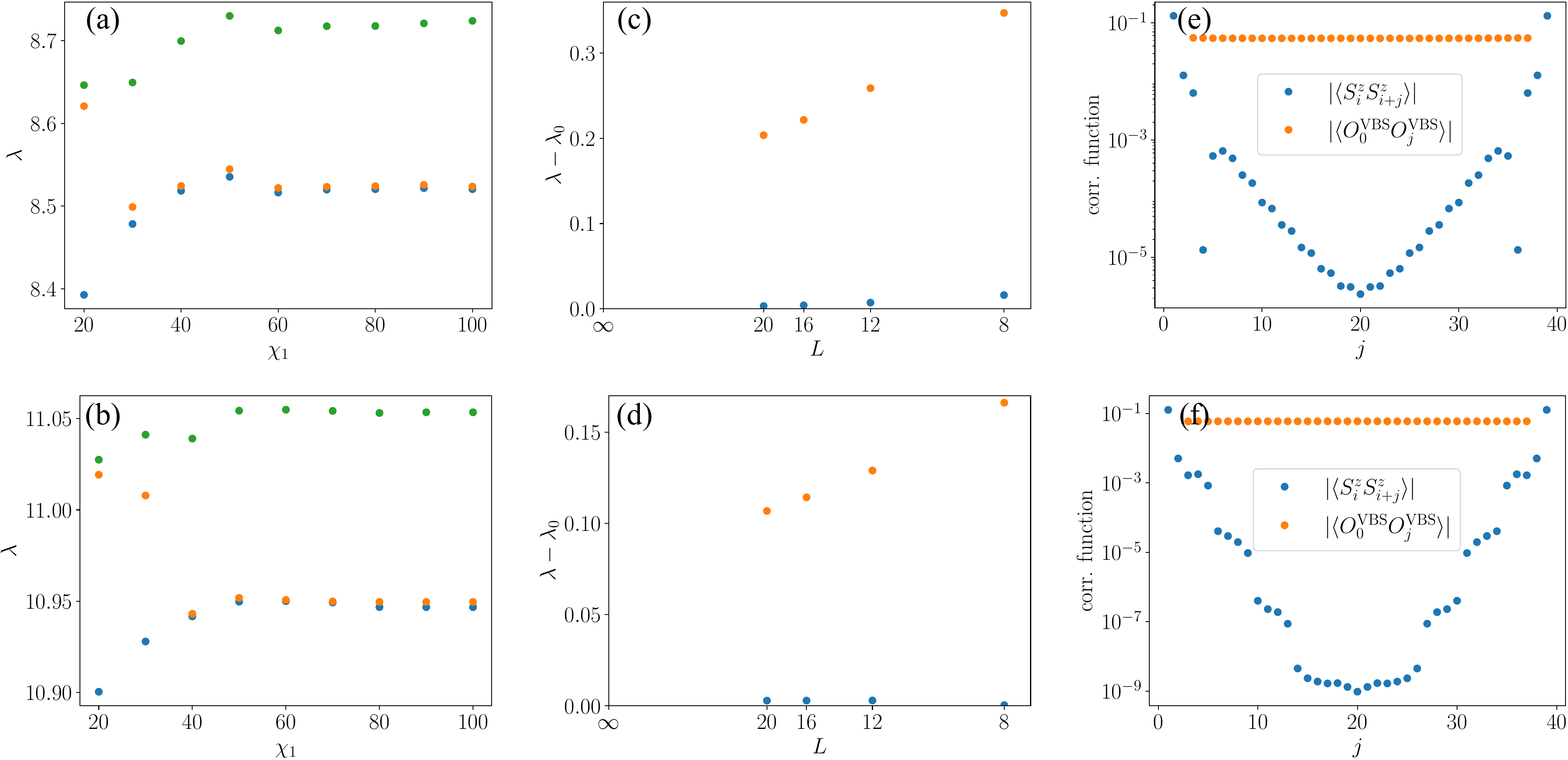}
    \caption{The numerical results of the MG ladders for (a, c, e) Model III and (b, d, f) Model IV. (a, b) The lowest three entanglement energies as a function of the bond dimension $\chi_1$ (see main text for details). Here $L=20$ and convergence is achieved around $\chi_1\simeq 80$. (c, d) The entanglement energy difference $\lambda_1-\lambda_0$ (blue) and $\lambda_2-\lambda_0$ (orange) of the entanglement Hamiltonian as a function of system size with $\chi_1=100$. (e, f) The correlation functions of spin $S^z$ (blue) and VBS order-parameter $O^\mathrm{VBS}$ (orange) for $L=40$ and $\chi_1=80$ as a function of the distance $j$. For the spin--spin correlation $|\langle S_i^z S_{i+j}^z\rangle|$ we have averaged over $i$. All the calculations are carried out with periodic boundary conditions.}
    \label{fig:mg}    
\end{figure*}

One can pinpoint the CFT more precisely by fitting the power-law exponent of the two-point correlation function. In anticipation of a logarithmic correction~\cite{Giamarchi2003}, we use three different fitting ans\"atze
\begin{align}
&\langle S_0^z S_j^z\rangle\propto\frac{1}{\Tilde{j}^\eta}, \label{eq:eta}\\
&\langle S_0^z S_j^z\rangle\propto\frac{(\ln(c\Tilde{j}))^\alpha}{\Tilde{j}^\eta}, \label{eq:alpha_eta} \\ 
&\langle S_0^z S_j^z\rangle\propto\frac{(\ln(c\Tilde{j}))^\alpha}{\Tilde{j}}, \label{eq:alpha}
\end{align}
with $\Tilde{j}=\sin(\pi j/L)$, see Fig.~\ref{fig:aklt2}. Without the logarithmic correction \eqref{eq:eta}, we get the critical exponent $\eta$ close to 0.9 [Fig.~\ref{fig:aklt2}(a, b)], while $\eta\simeq1.0$ when such corrections are allowed [Fig.~\ref{fig:aklt2}(c, d)]. In the third ansatz \eqref{eq:alpha}, we fix $\eta=1$ and fit against the exponent of the logarithmic correction $\alpha$ alone [Fig.~\ref{fig:aklt2}(e, f)]. We find that both \eqref{eq:alpha_eta} and \eqref{eq:alpha} give excellent fitting of the numerical data, in full consistence with $\eta=1$. Together with the symmetry, this strongly suggests that the low energy theory is one of SU$(2)_1$ Wess--Zumino--Witten CFT \cite{Francesco1997}.

An important feature of the bipartite entanglement spectrum of a pure state is that it is identical for both parts (up to zeros that go to infinity after taking a logarithm). In the case of spin-1/2 coupled to spin-3/2, this guarantees that the spin-3/2 entanglement Hamiltonian has the identical spectrum as the spin-1/2, and therefore the same central charge $c=1$. Historically, the universality class of the spin-3/2 Heisenberg chain had caused some confusion before it became clear that it is the same as its spin-1/2 counterpart \cite{Affleck1986b,Affleck1987b,Affleck1989,Hallberg1996}. In the current setup, on the other hand, a knowledge of the spin-1/2 automatically results in that of the spin-3/2 system, where the MPDO of the latter can be similarly obtained.

\section{Gapped degenerate entanglement Hamiltonian of the MG ladders}
We now turn to the decohered MG ladders, where one expects a gapped, two-fold degenerate entanglement Hamiltonian. Indeed, one expects a spontaneous symmetry breaking (SSB) behavior, where the translational symmetry $\mathbb{Z}_L$ is broken to $\mathbb{Z}_{L/2}$\footnote{We take the system size $L$ to be even for simplicity.}. Again, we consider one proposed in Ref.~\cite{zhou2023} with a spin-$1/2$ chain ($a$) as the system coupled to spin-$3/2$ modes ($b$) as the bath (Model IV) and another with a spin-$1/2$ bath (Model III). The Hamiltonians of Model III and IV reads
\begin{align}
{H}_\mathrm{III}=\sum_{i=1}^L ~~&J_1({\bf {S}}_{i,a}\cdot {\bf S}_{i+1,a}+\frac{1}{2}{\bf {S}}_{i,a}\cdot {\bf {S}}_{i+2,a})\nonumber\\
+ ~&J_2 {\bf {S}}_{i,a}\cdot {\bf {S}}_{i,b},\\
{H}_\mathrm{IV}=\sum_{i=1}^L ~~&J_1({\bf {S}}_{i,a}\cdot {\bf S}_{i+1,a}+\frac{1}{2}{\bf {S}}_{i,a}\cdot {\bf {S}}_{i+2,a})\nonumber\\
+ ~&J_2 {\bf {S}}_{i,a}\cdot {\bf {S}}_{i,b}+D({S}_{i,a}^z+{S}_{i,b}^z)^2,         
\end{align}
where a spin-$1/2$ MG chain ($a$) \cite{Majumdar1969} is coupled to a spin-$1/2$ (spin-$3/2$ respectively) bath ($b$) such that the total system is trivially gapped and we take $J_1=J_2=D=1$. In Model III, the rung coupling favors a singlet, while an $|S=1,m=0\rangle$ state is preferred on each rung in Model IV. Note that Model IV has O(2) instead of SO(3) symmetry which already suffices for LSM \cite{Affleck1986}. Due to the lack of an exact solution to the ladder problem, one has to resort to DMRG twice --- once to obtain the (physical) ground state and again to find the low energy states of the entanglement Hamiltonian. In the second step, we have contracted the MPS on each site to obtain an MPDO. To obtain an MPDO with a reasonable bond dimension, we have to use a limited bond dimension $\chi_1$ in the first step, and considerable truncation errors are inevitable compared to the usual DMRG precision. The bond dimension in the second step is fixed at $\chi_2=100$, sufficient for a gapped spin-$1/2$ system with $L$ a few decades. To ensure the reliability of the results, we change the bond dimension $\chi_1$ and see how the entanglement spectrum follows. We find that convergence is quickly reached around $\chi_1\simeq 80$, see Fig.~\ref{fig:mg}(a, b). Again, we can attribute the convergence to the quasi-localness of $K$. With this, we calculate the low entanglement energy spectrum for different system sizes and the results are shown in Fig.~\ref{fig:mg}(c, d). Both the two-fold degeneracy of the ground states and the finite gap above them have already manifested with $L$ up to 20.

It is of interest to verify the order-parameter given the expected SSB nature. The translational-symmetry-breaking valence-bond-solid (VBS) order-parameter can be chosen as
\begin{equation}
O_i^\mathrm{VBS}=S_{i-1}^z S_{i}^z-S_i^z S_{i+1}^z,
\end{equation}
long-range correlation of which is corroborated in Fig.~\ref{fig:mg}(e, f). This is contrasted with the exponential decay of the spin--spin correlation function.

\section{Conclusions}
Compared to sub-dimensional entanglement properties, many key questions concerning extensive entanglement remain to be answered. In this work, motivated by the recent progress of the open system LSM theorem, we carry out extensive numerical investigations of extensive entanglement properties in quantum spin ladders. While the total system is trivially gapped and its ground state has SRE, symmetry constraints dictate that the entanglement spectrum must be either gapless or SSB, or equivalently, the ground state of $K$ has LRE. Apart from fully corroborating the analytical proposals in both cases of gaplessness and SSB, our numerical results provide many more details on the universality properties using CFT diagnoses in the former case. Given the exact nature of the AKLT MPDO, we expect this model, or its higher spin generalizations, to assume a paradigmatic role in future research on extensive entanglement. Regarding the MG ladders, our results showcase the capacity of approximate MPDOs from MPS to capture the SSB characteristics. We expect this work to lay the foundations for more numerical works on extensive entanglement. In the scenarios where the two 1d chains can be regarded as boundaries of 2d systems, our results also shed light on the more conventional sub-dimensional entanglement properties.  On the analytical side, some pressing questions immediately follow from our results, including whether other CFTs can be realized in similar setups, and whether the quasi-local nature of the extensive entanglement Hamiltonians allows phases beyond the strictly local case \cite{Li2023}. Much remains to be asked and answered in this largely uncharted territory.\\

\noindent
{\bf Acknowledgements}
We thank Yingfei Gu and Hui Zhai for helpful discussions and collaboration on the preceding work. The DMRG calculations are performed using the ITensor library (v0.3) \cite{ITensor}.\\

\noindent
{\bf Funding information}
This work is supported by the Innovation Program for Quantum Science and Technology 2021ZD0302005, the Beijing Outstanding Young Scholar Program, the XPLORER Prize, and China Postdoctoral Science Foundation (Grant No. 2022M711868). C.L. is also supported by the Chinese International Postdoctoral Exchange Fellowship Program and the Shuimu Tsinghua Scholar Program at Tsinghua University. \\

\noindent
{\bf Availability of data and materials}
All data underlying the results are available from the authors upon reasonable request.\\

\noindent
{\bf Competing interests}
The authors declare no competing interests.\\

\noindent
{\bf Author contributions}
CL did the theoretical and numerical work with the input from XL and YNZ. All authors participated in the analysis of the results. CL wrote the manuscript with the input from all authors. All authors have read and approved the manuscript.

\bibliography{main.bib}

\end{document}